\documentstyle[prb,aps,multicol,epsfig]{revtex}
\tighten
\newcommand{\beq}{\begin{equation}}
\newcommand{\eeq}{\end{equation}}
\newcommand{\bea}{\begin{eqnarray}}
\newcommand{\eea}{\end{eqnarray}}

\begin{document}
\draft

\title{Adsorption of charged particles on an oppositely charged surface:
  Oscillating inversion of charge}

\author{Toan T. Nguyen and Boris I. Shklovskii}

\address{Department of Physics, University of Minnesota, 116 Church
  St. Southeast, Minneapolis, Minnesota 55455} 

\maketitle

\begin{abstract}
Adsorption of multivalent counterions on the charged surface of a
macroion is known to lead to inversion of the macroion charge due to
the strong lateral correlations of counterions. We consider a
nontrivial role of the excluded volume of counterions on this effect.
It is shown analytically that when the bare charge of macroion
increases, its net charge including the adsorbed counterions
oscillates with the number of their layers. Charge inversion vanishes
every time the top layer of counterions is completely full and becomes
incompressible. These oscillations of charge inversion are confirmed
by Monte-Carlo simulations. Another version of this phenomenon is
studied for a metallic electrode screened by multivalent counterions
when potential of the electrode is controlled instead of its charge.
In this case, oscillations of the compressibility and charge inversion
lead to oscillations of capacitance of this electrode with the number
of adsorbed layers of multivalent counterions.
\end{abstract}


\begin{multicols}{2}

\section {Introduction}
Adsorption of charged particles on the surface of an oppositely
charged macroion is an important problem with broad applications in
many areas of science.  In a water solution, double helix DNA, actin,
charged colloids, charged membranes or any charged interfaces can play
the role of the macroion. Charged particles, or counterions, in
solution can be ions, small colloids, charged micelles, short or long
polyelectrolytes.  Mean field theories based on the Poisson-Boltzmann
equation and its linearized version, the Debye-H\"{u}ckel equation,
have been used to study such screening problem.  There are, however,
several new phenomena in solutions containing multivalent counterions
which cannot be explained using standard mean-field theories.  The
most notorious of all is probably the charge inversion, a counter
intuitive phenomenon in which a macroion strongly binds so many
counterions that its net charge changes sign.  This can be thought of,
theoretically, as overscreening. It cannot be explained using
Poisson-Boltzmann theory because, in a mean field theory, screening
compensates at any finite distance only a part of the macroion charge.
Charge inversion recently has attracted significant
attention.\cite{Roland,Perel,Shklov99,Pincus,Joanny0,Sense,Joanny1,Joanny2,Dubin,Linse,Shklov992,Nato,Holm,Rubinstein,Stoll,Andelman,Khokhlov,Matochiko,Netz}

\begin{figure}
\epsfxsize=5.5cm \centerline{\epsfbox{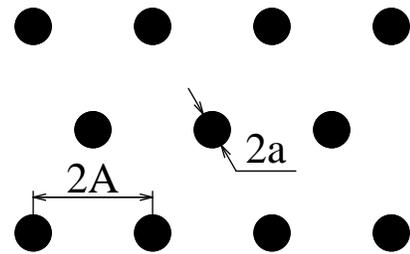}}
\caption{The two-dimensional strongly correlated liquid (SCL) of $Z$-ions on
  the macroion surface with the short range order of a
  Wigner crystal. Black disks are $Z$-ions with
  radius $a$. The lattice constant of the crystal is
  $2A=\sqrt{2/\sqrt{3}n}$.}
\label{WCfig}
\end{figure}

\begin{figure}
\epsfxsize=7.5cm \centerline{\epsfbox{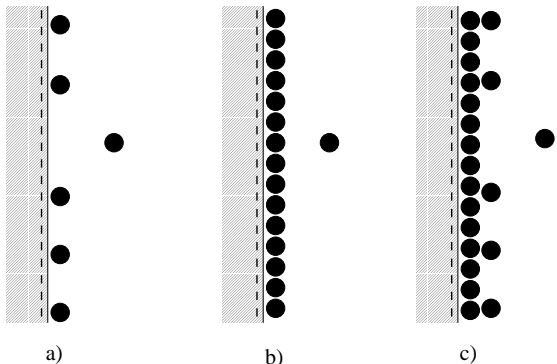}}
\caption{The origin of oscillations of charge inversion.
  a) The adsorbed $Z$-ion layer is only partially filled.
  A new $Z$-ion pushes other ions aside to create a correlation
  hole for itself. The $Z$-ion is closer to the negative
  surface charge of this correlation hole than to the positive
  neighbouring $Z$-ions and therefore is attracted to the surface.
  b) The layer is filled and  correlation hole disappears.
  c) More than one layer is filled.
  The correlation hole of the approaching $Z$-ion is created in the
  top layer only.}
\label{layer}
\end{figure}

It was shown\cite{Perel,Shklov99,Joanny1,Shklov992,Nato,Holm,Rubinstein,Netz} that
charge inversion is driven by the counterion correlations which are
ignored in mean field theories\cite{metallization}.  
When multivalent counterions condense
on the surface of a macroion to screen its charge, due to their strong
lateral repulsion they form a two-dimensional strongly correlated
liquid. To see when correlation is important in this liquid, one first
defines a dimensionless parameter $\Gamma$ which measures the strength
of the interaction energy between counterions in units of the thermal energy:
$\Gamma = (Ze)^2 (\pi n)^{1/2}/D k_BT$, where $D\simeq 80$ is the dielectric
constant of water, $Ze$ is the charge of a counterion (for
convenience, we call it a $Z$-ion) and $n$ is the two-dimensional
concentration of $Z$-ions condensed on the macroion surface.  Without
loss of generality, we assume through out this paper that $Ze$ is
positive and the macroion is negatively charged with the surface
charge density $-\sigma$.  For a strongly charged macroion with
surface charge density $-\sigma\sim -1e/nm^2$, $\Gamma \simeq 1.2,\ 
3.5,\ 6.4$ and $9.9$ at $Z=1,\ 2,\ 3$ and $4$.  Thus, for $Z \ge 3$,
$\Gamma$ is a large parameter of the theory and the two-dimensional
system of $Z$-ions is a strongly correlated liquid (SCL).  It has
short range order very similar to a two-dimensional Wigner crystal
(See Fig.  (\ref{WCfig})).  At the same time, large parameter $\Gamma$
also guarantees the layer of $Z$-ions at the macroion surface is
effectively two-dimensional.  Indeed, each $Z$-ion, in the uniform
field $2\pi\sigma$ of the macroion surface, moves within a distance
$Dk_BT/2\pi\sigma Ze 
\simeq 0.52 A/\Gamma$ from the macroion surface.  At large $\Gamma$,
this distance is much smaller than average distance $2A$ between
$Z$-ions in the direction parallel to the macroion surface. This makes
the $Z$-ion layer effectively two-dimensional.
The correlation physics explains charge inversion as 
follows\cite{Shklov99,Shklov992,Nato}. When a new
$Z$-ion comes to the macroion which is already neutralized by the
$Z$-ion layer, it pushes other ions aside and creates a negative
background charge for itself. In other words, it creates an oppositely
charged image (or a correlation hole) in the $Z$-ion layer (Fig.
\ref{layer}a).  Due to the attraction from this image, the $Z$-ion
sticks to the surface and overcharges it.  The overcharging saturates
when the positive inverted net charge of the macroion is large enough
to counterbalance the above mentioned attractive force. In a more quantitative
language, correlation
induced attraction to the $Z$-ion layer can be explained as a
result of the negative chemical potential $\mu_{SCL}$ of SCL, whose magnitude is
much larger than $k_BT$ ($|\mu_{SCL}|\sim \Gamma k_BT \gg k_BT$).

\begin{figure}
\epsfxsize=8cm \centerline{\epsfbox{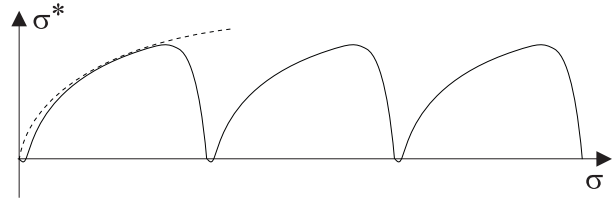}}
\caption{Schematic plot of the macroion net surface charge density
  $\sigma^*$ as a function of the absolute value of its bare charge
  density $\sigma$. Every time a layer of $Z$-ions is almost full,
  charge inversion disappears and then recovers when the next $Z$-ion
  layer builds up. The dashed line which is described by Eq.
  (\ref{maxq}) is the maximum charge inversion one obtains if the
  effect of excluded volume is neglected ($a=0$). }
\label{qheplot}
\end{figure}

In this paper, we would like to investigate the problem of charge
inversion further by considering the effect of the excluded volumes of
$Z$-ions.  This effect was studied in Poisson-Boltzmann
approach\cite{Borukhov}. Here we show that in the correlation theory,
this effect can lead to qualitatively new consequences.  When the
charge of the macroion increases, $A$ decreases and $\Gamma$
increases, the correlation between $Z$-ions becomes stronger. In other
words, $|\mu_{SCL}|$ increases and charge inversion is stronger.
However, the excluded volume has a dramatic consequence when $A
\rightarrow a$: the pressure of the SCL liquid diverges and it becomes
effectively incompressible.
In this case, the correlation image and charge inversion disappears
(Fig. \ref{layer}b). If the macroion charge increases further,
however, a second layer of $Z$-ions is created and charge inversion
recovers (Fig. \ref{layer}c).
Here everything is repeated again but with the role of $\sigma$
played by the net charge of the macroion and the first full layer.
Consequently, if one plots the net charge of the macroion versus its
bare charge, one sees an oscillation of charge inversion: every time a
new layer of $Z$-ions forms on the macroion surface, charge
inversion appears and steadily increases with increasing macroion
charge then drops quite abruptly when the layer starts to become full
(see Fig. \ref{qheplot}).  As a result of these oscillations, the
maximum value of the inverted net charge is limited, no matter how large
the charge of the macroion is.  To our knowledge, this is the first
time this periodical behaviour is suggested.

This paper is organized as follows. In Sec. II, we derive an
analytical formula for the chemical potential of the $Z$-ions
condensed on the macroion surface taking into account the excluded
volume effect. This chemical potential is then used to calculate the
inverted charge of the macroion. In Sec. III, a Monte Carlo simulation
of a small system is used to verify the analytical predictions of Sec.
II. In Sec. IV, we consider the role of charge inversion in the case
of screening of a metallic electrode, whose potential is tuned
instead of surface charge density $\sigma$. We show that, in this
case, correlations lead to the oscillations of the macroion
capacitance.

\section{Chemical potential of $Z$-ions with hard-core repulsion.}

To understand the electrostatics of the system, let us start with a
two-dimensional system of point charges ($a=0$) with concentration $n$
on a neutralizing background charge. Conventionally, this system is
referred to as a one component plasma. Excluded volume effects will be
considered later.  At large parameter $\Gamma$, the plasma is a
strongly correlated liquid (SCL) whose short range order is very
similar to a Wigner crystal (WC).  Specifically, the energy per
$Z$-ion $\varepsilon_{SCL}(n)$ of a SCL can be very well approximated
by that of a WC, which in turn can be approximated by the energy of a
Wigner-Seitz cell because quadrupole-quadrupole interactions between
neutral Wigner-Seitz cells are very small. This gives
\begin{equation}
\varepsilon_{SCL}(n)\sim \varepsilon_{WC}(n) 
        \simeq-1.10\frac{Z^2e^2}{AD}~~.
\end{equation}
A more accurate calculation\cite{Mara} gives 
$\varepsilon_{WC}(n)$ a slightly higher value:
\begin{equation}
\varepsilon_{WC}(n) \simeq -1.06Z^2e^2/AD~~.
\end{equation}
Knowing $\varepsilon_{WC}(n)$, the chemical potential of the WC
can be easily calculated:
\begin{equation}
\mu_{WC}(n)=\frac{\partial [n\varepsilon_{WC}(n)]}{\partial n}
\simeq -1.57 \frac{Z^2e^2}{AD}=-1.65 \Gamma k_BT
\end{equation}
Thus, $\mu_{WC}(n)$ is negative and much larger than $k_BT$. It acts
as an additional attraction to the macroion surface which in turn
leads to the charge inversion effect.  The temperature correction to
$\mu_{WC}(n)$, or in other words, the difference between
$\mu_{SCL}(n)$ and $\mu_{WC}(n)$, has been calculated numerically
\cite{Totsuji} for a wide range of the parameter $\Gamma$ and indeed
is very small and can be neglected. For example, it is equal to 11\%
at $\Gamma=5$ and 5\% at $\Gamma=15$.

Now let us return to our problem of finite size $Z$-ions.  As the
detail derivation in the appendix shows, the chemical potential of a
$Z$-ion with hard-core repulsion can be written as the sum
\begin{equation}
\label{mu}
\mu(n)=\mu_0(n)+\mu_{WC}(n)~~,
\end{equation}
where $\mu_0$ is the chemical potential of a system of neutral hard
discs of radius $a$ at the same concentration $n$. In writing down Eq.
(\ref{mu}), the electrostatic contribution to $\mu(n)$ has been
approximated by $\mu_{WC}(n)$, in exactly the same way $\mu_{SCL}(n)$
was approximated by $\mu_{WC}(n)$ above.  The temperature correction
to $\mu_{WC}(n)$, in this case, is even smaller than that for the SCL
of point charges because due to the hard-core repulsion, two $Z$-ions
are never at a distance less than $2a$ from each other.  Furthermore,
this correction decreases when $A$ decreases making the approximation
by $\mu_{WC}(n)$ increasingly better and asymptotically exact when
$A\rightarrow a$.

To proceed further, one needs to calculate the chemical potential
$\mu_0(n)$.  The hard disc and hard sphere systems have been
extensively studied both analytically and numerically providing
significant insights into solid-liquid phase transition and the
melting process (See Ref. \onlinecite{Ziman} and references therein).
In this paper, to emphasize the effects of excluded volume, we would
like to repeat here simple calculations using the free volume
approximation which is asymptotically exact near the close packing
limit.

\begin{figure}
\epsfxsize=5cm \centerline{\epsfbox{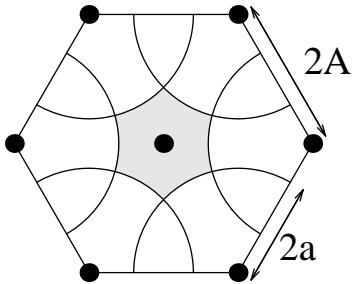}}
\caption{The center of the central $Z$-ion is confined in the gray
  area by the hard core repulsion from its six nearest neighbours on
  the hexagonal lattice (the black dots). Near close packing, this
  area is significantly smaller than the area of the hexagon itself.
  }
\label{FV}
\end{figure}

To calculate $\mu_0$, one notices that because of the hard-core
repulsion, the area $s$ available per each disc is smaller than the
area $2\sqrt{3} A^2$ of a Wigner-Seitz cell (See Fig. \ref{FV}).
A simple calculation shows that near the close packing limit,
$s=\alpha(A-a)^2 \ll 2\sqrt{3} A^2$, where $\alpha \simeq 4.62$.  To
calculate the free energy of the quasi two-dimensional layer of
$Z$-ions, one needs to add a third out-of-surface dimension.  The
typical distance $x$ of the out-of-surface motion of a $Z$-ion is the
distance at which the energy cost for moving the $Z$-ion away from the
plane is of the order of $k_BT$.  Solving $2\pi\sigma Zex/D\simeq
k_BT$ gives $x=Dk_BT/2\pi\sigma Ze=\Lambda$ which coincides with the
Gouy-Chapman distance (note that this estimation of $x$ is consistent
because, for $\Gamma \gg 1$, $\Lambda \ll A$, all other $Z$-ions are
far apart and we can use the planar electric field $2\pi\sigma/D$ for
the calculation of $x$.)  Now, one can estimate the free energy of the
hard disc system as the free energy of an ideal gas with concentration
$1/s\Lambda$:
\begin{equation}
F_0=-Nk_BT\left(\ln\frac{s\Lambda}{v_0}+1\right)~,
\label{F0}
\end{equation}
where $N$ is the number of adsorbed $Z$-ions and $v_0$ is the normalizing 
volume which can be thought of as the
volume of one water molecule. The exact value of $v_0$ is not
important because it drops out of all final results.

The chemical potential of adsorbed $Z$-ions is, correspondingly:
\begin{eqnarray}
\mu_0 = \frac{\partial F_0}{\partial N}&=&
        k_BT\left(\ln\frac{v_0}{s\Lambda}-1\right)-
        k_BT\frac{N}{s}\frac{\partial s}{\partial N} \nonumber \\
        &\simeq&
        k_BT\left(\ln\frac{v_0}{\alpha(A-a)^2\Lambda}-1\right)+
        k_BT\frac{A}{A-a} ~~.
\label{mu0}
\end{eqnarray}
From Eq. (\ref{mu0}), one can see that when $A - a \ll a$,
this chemical potential is positive and very large what negates any
gain from the electrostatic part $\mu_{WC}$.  Charge inversion
therefore disappears in this limit and will not recover until the
second layer builds up when the macroion charge grows.

Using the correlation hole concept mentioned in the 
introduction, one can easily understand why charge inversion
disappears.  The second term in Eq. (\ref{mu0}), which is the dominant
term when $A$ approaches $a$, is nothing but $p_0/n$, where
$p_0=-\partial F_0/\partial S=nk_BTA/(A-a)$ is the two-dimensional 
pressure of the
hard disc system near the close packing limit, $S$ is the total area
of the macroion surface.  The divergence of the pressure leads to the
zero compressibility of the $Z$-ion layer when $A\rightarrow a$. When
a new $Z$-ion approaches this layer, it cannot put other $Z$-ions
aside to create its correlation hole to attract to it (Fig.
\ref{layer}b).  Therefore, charge inversion disappears.

It should be noted that, even though Eq. (\ref{F0}) is expected to be
asymptotically correct only near the close packing limit
($A\rightarrow a$), where the hard disc system is known to be in the
solid phase, it is actually quite good as an approximation in the
whole range of $A$.  Indeed, Monte-Carlo simulations and molecular
dynamics calculations (See Ref. \onlinecite{Wood}) show that the
pressure obtained using the free volume approximation gives an
excellent result for the solid phase of the hard disc system and a
good result for the liquid phase.  The biggest discrepancy occurs
around the liquid-solid phase transition is only about 20\%.
Combining with the fact that, for large $\Gamma$, the electrostatic
contribution $\mu_{WC}$ is the dominant term of $\mu$ everywhere
except near close packing, we can reliably use Eq. (\ref{mu}) for the
$Z$-ion chemical potential.

Let us now calculate the net charge of the macroion which is the total
charge of the macroion and the $Z$-ion layer on its surface. To do so,
one has to balance the electro-chemical potential of the $Z$-ions at
the macroion surface with the electro-chemical potential $\mu_b$ of
$Z$-ions in the bulk solution.  If we assume the concentration of the
$Z$-ions in the bulk solution is $c$, this balance in
electro-chemical potential reads:
\begin{equation}
\mu + Ze\psi(0) = \mu_b \simeq k_BT\ln(cv_0),
\label{equi1}
\end{equation}
where $\psi(0)=2\pi\sigma^* r_s$ is the averaged macroscopic
electrostatic potential at the macroion surface due to the net charge
density $\sigma^*=-\sigma+Ze/2\sqrt{3} A^2$ of the macroion surface with
$Z$-ions condensed on it, $r_s$ is the screening length of the
solution. On the right side of Eq. (\ref{equi1}), the bulk solution
(where the electrostatic potential vanishes) is assumed to be ideal so that
$\mu_b$ is approximated by the ideal gas chemical potential
$k_BT\ln(cv_0)$.

Substituting Eq. (\ref{mu}) and (\ref{mu0}) into Eq. (\ref{equi1}),
one gets:
\begin{equation}
Ze\psi(0)=|\mu_{WC}|-k_BT\ln\frac{1}{\alpha(A-a)^2\Lambda c}-
        k_BT\frac{a}{A-a}~.
\label{equi2}
\end{equation}
Because $\mu_{WC}$ is negative and much larger than $k_BT$ in absolute
value, one can see from Eq. (\ref{equi2}) that, if $A$ is not very
close to $a$, the last two terms on the right hand side can be
neglected. In this case, $\psi(0)$ and therefore $\sigma^*$ are
positive. Recalling that the bare charge density $-\sigma$ of the
macroion is negative, we see a charge inversion effect: the amount of
$Z$-ions adsorbed by the macroion surface is larger than needed to
neutralize its charge.  Eq.  (\ref{equi2}) clearly shows that this
effect is driven by correlations since the correlation part
$|\mu_{WC}|$ of the chemical potential is the only positive term in
the right hand side.

It is interesting to mention the limiting case of Eq. (\ref{equi2})
in which $Z$-ions are point charges ($a=0$). At high enough bulk
concentration $c$ or low enough temperature $T$, Equation
(\ref{equi2}) 
then gives
$Ze\psi(0)=|\mu_{WC}|$. For a spherical macroion, its surface
potential is $\psi(0)=Q^*/r_M D$
($r_M$ is the macroion radius, $Q^*$ is the macroion net charge). This
gives\cite{Shklov99}:
\begin{equation}
Q^*=0.84\sqrt{QZe}~~,
\label{maxq}
\end{equation}
where $Q$ is the macroion bare charge. This is the maximum possible
inverted charge of a spherical ion. Surprisingly, this charge does not depend
on the macroion radius. This function $Q^*(Q)$ is plotted in Fig.
\ref{qheplot} by the dashed line.

For $Z$-ions with finite radius $a$, when $A$ approaches $a$ the last
two terms in Eq. (\ref{equi2}) increase substantially and reduce
charge inversion. When a near complete layer of $Z$-ion is formed
($A-a \ll a$), these terms eventually become bigger than $|\mu_{WC}|$.
When this happens, $\sigma^*$ becomes negative and charge inversion
disappears.  If one increases the macroion charge further, the second
layer of $Z$-ions appears on the surface of the macroion and the
charge inversion recovers. Now, one can still use Eq. (\ref{equi2}) to
calculate charge inversion provided the bare charge of the macroion is
replaced by the net charge of the macroion with the first layer of
$Z$-ion $\sigma_1=-\sigma+Ze/2\sqrt{3} a^2$.  Similarly, as the
macroion ion charge increases further one obtains a periodic behaviour
of charge inversion with respect to the increase of macroion charge:
charge inversion disappears every time a layer of $Z$-ions is near
full and recovers when new layer appears.
This oscillating behaviour is shown in Fig. \ref{qheplot} where the
solution to Eq.  (\ref{equi2}) is plotted. In Fig. \ref{qheplot},
there are small regions near the beginning of each layer where the net
charge density of the macroion with full layers is so small that
$\Gamma < 1$. In such a region, Eq.  (\ref{equi2}) cannot be used to
describe screening of the macroion surface. Instead, when $\Gamma <1$,
correlations can be neglected and one should use the standard
Poisson-Boltzmann mean-field screening theory. Hence, in this region
the net charge is the same as the bare macroion charge ($\sigma^* <
0)$.  As $\sigma$ increases, however, $\Gamma$ soon reaches unity and
charge inversion takes over.

It should also be noted here that, in all above calculations,
the dielectric constant of the macroion was implicitly assumed to be
equal to the dielectric constant $D$ of water solution so that all
$Z$-ions lie on the macroion surface. In general, the dielectric
constant of a macroion is smaller than water so that a $Z$-ion
approaching the surface sees an additional repulsive force from its
electrostatic image in the macroion due to the discontinuity in the
dielectric constant at the macroion-water interface. This additional
image pushes the $Z$-ion off at a distance from the surface and
reduces the attractive force created by the correlation hole in the
adsorbed $Z$-ion layer calculated above. However, all the oscillating
charge inversion picture mentioned above remains valid qualitatively.
Indeed, it is shown in Ref. \onlinecite{Shklov992} that the additional
image charge lifts each $Z$-ion off the surface by a distance of the
order of $A/4$ and reduces the correlation energy approximately by
half, thus $\mu_{WC}$ remains highly negative and
charge inversion is preserved. 

\section{Monte-Carlo simulation}

To demonstrate the oscillating behaviour of charge inversion, we carry
out Monte-Carlo simulations of a small system of a macroion with
$Z$-ions.  The $Z$-ions are modeled as charged spheres with charge
$Z=4$ and radius $a=0.9l_B$, where the Bjerrum length $l_B =
e^2/Dk_BT$. The system is simulated at room temperature ($298^{\circ}$K)
so that $l_B \simeq 7.2$\AA.
The macroion has radius
$r_M=3.5l_B$. Thus, a full (hexagonal close-packing) first layer of
$Z$-ions would contain $4\pi r_M^2/2\sqrt{3}a^2\sim 86$ $Z$-ions if
one does not take into account defects caused by the finite
curvature of the macroion surface (one cannot put a perfect crystal on
the surface of a sphere). The actual number of $Z$-ion at full filling
should be smaller.  In our Monte-Carlo simulation, in part also due to the
divergence of pressure mentioned in the previous section, the number
of $Z$-ions in the first layer never exceeds 75.  

\begin{figure}
\epsfxsize=4cm \centerline{\epsfbox{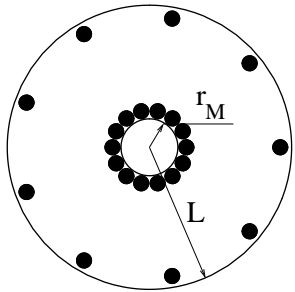}}
\caption{Schematic drawing of the Monte-Carlo cell. The macroion is
  fixed at the center. Some $Z$-ions condense on the macroion
  and overscreen it. The rest of the $Z$-ions spread out over the outer
  boundary of the system.}
\label{schem}
\end{figure}

The macroion charge
$-Q$ is varied from 20 to 550 so that at maximum there are two layers
of $Z$-ions on the macroion surface. 
Energies of interaction between two $Z$-ions and
between each $Z$-ion and the macroion
were calculated according to unscreened Coulomb law 
$(Ze)^2/Dr_{ij}$ and
$- ZQe^2/Dr_{Mi}$, where $r_{ij}$ is the distances between 
centers of the
two $Z$-ions numbered $i$ and $j$, and
$r_{Mi}$ is the distances between the centers of the $Z$-ion
numbered $i$ and the macroion. The net macroion charge
$Q^*=-Q+NZe$ is determined from the result of the simulation where $N$
is the number of $Z$-ions adsorbed on the macroion. A $Z$-ion is
considered adsorbed if its center is found within a distance $3a$ from
the top full layer of the $Z$-ion (or from the macroion surface if
there's only one partially filled layer).  The whole system is
positioned inside a hard spherical shell with large radius $L=17.4l_B$
with the macroion fixed at the center of the shell (see Fig.
\ref{schem}).
In our simulated system, the total charge of the macroion and all the
$Z$-ions is $220e$, so that the total charge of $Z$-ions is larger
than the macroion charge.  This is needed to observe charge inversion.
In practical situation, the excess charge of $Z$-ions is neutralized
by monovalent ions at the distance of the order of the screening
length $r_s$ from the macroion surface.  In our simulation, these excess
$Z$-ions are neutralized at the external hard spherical shell instead
(it would be too time consuming to simulate the monovalent ions). This
difference, however, produces no significant impact on the charge
inversion because of the short range nature of correlations.
Indeed, it is shown in Ref. \onlinecite{Shklov99} that if $r_s$
is larger than radius of the macroion,
charge inversion only weakly depends on $r_s$.
This is because the capacitance of the macroion screened at the distance
much larger than its radius only weakly depends on $r_s$.
In the same way, charge inversion of the macroion 
in the shell is almost independent
on the shell radius when it is much larger then macroion radius.
This is because the capacitance of macroion-shell capacitor is
almost constant, while it is charged by the negative chemical
potential of $Z$-ion which is determined by interactions
at the distances of the order
of the size correlation hole, $A$, from the macroion surface.

For most of the Monte-Carlo attempted moves, the new position of a
randomly chosen $Z$-ion is chosen with uniform distribution in a cube
with size $l_B$ centered at the old $Z$-ion position.  However, once
in every 100 attempts, a longer attempted move is made:
If the chosen $Z$-ion is found in the vicinity
of the macroion ($r_M < r < r_M+2a$, where $r$ is the distance from
the $Z$-ion to the macroion center), it is attempted to move to the
vicinity of the external shell ($L-3.5a < r < L$).  Vice versa, if the
chosen $Z$-ion is found in the vicinity of the external shell, it is
attempted to move to the vicinity of the macroion. This is done to
quickly equilibrate the system and to overcome
the huge Coulomb barrier which results from the combination of short
range attraction of a $Z$-ion to its image in the SCL and the
repulsion from the 
inverted macroion charge. The usual Metropolis algorithm is used to
reject or accept a move.  Typically, in a simulation run, the number
of attempted moves per $Z$-ion is $2\times 10^6$, of which the last
$1\times 10^6$ moves are used for averaging of $Q^*$.

\begin{figure}
\epsfxsize=5cm \centerline{\epsfbox{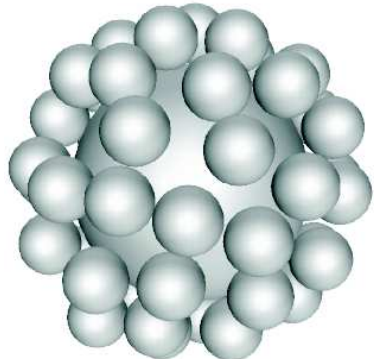}}
\caption{A snapshot of the macroion with the layer of
  $Z$-ions condensed on it. The macroion charge is $-220e$. All
  $Z$-ions found within a distance less than $3a$ from the macroion
  are shown.  In this snapshot, there are 58 of them making the
  macroion net charge $4e\times 58-220e=12e$.}
\label{snap}
\end{figure}

\begin{figure}
\epsfxsize=7cm \centerline{\epsfbox{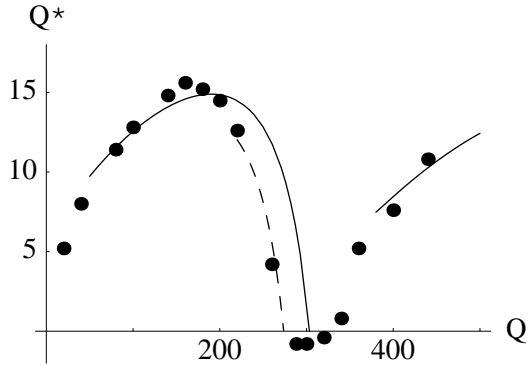}}
\caption{The net macroion charge $Q^*$ vs. its bare charge $Q$
  as the result of Monte-Carlo simulations (the solid circles). The
  solid line is the theoretical prediction using Eq. (\ref{equi2L}).
  We do not extend theoretical predictions for $Q<50$ and for the
  beginning of the second layer because, there, predictions of Eq.
  (\ref{equi2L}) loose their accuracy (see the text). The dash line
  corresponds to the corrected theoretical prediction near the full
  filling, which takes into account lattice defects.}
\label{mcresult}
\end{figure}

A snapshot of the complex of the macroion and the $Z$-ions condensed
on it is shown in Fig. \ref{snap} for the case the macroion charge is
$-220e$. The net macroion charge in this case is $12e$ (the macroion
is overscreened).
The simulation results are shown in Fig. \ref{mcresult},
where the macroion net charge $Q^*$ is plotted against the macroion
bare charge $Q$ together with our theoretical prediction. 
As we expected, the charge inversion disappears ($Q^*$ goes to zero)
near complete filling. After that, the second layer of condensed
$Z$-ions starts to form and the macroion is undercharged initially.
When $Q$ increases further so that more $Z$-ions come to the second
layer and correlations become important, charge inversion soon
recovers and increases with $Q$.
In order to compare our theory with simulation results some
modifications are necessary. Because of the presence of the hard shell
which confines the $Z$-ions, Eq. (\ref{equi1}) should read:
\begin{equation}
\mu+Ze \psi(0)=k_BT\ln(c_Lv_0)+Ze\psi(L)~,
\end{equation}
where in the right hand side, the bulk concentration $c$ is replaced
by the concentration $c_L$ at distance $L$. Because $c_L$ is small, we
still assume ideal solution neglecting $Z$-ion correlations (so that
the chemical potential of $Z$-ions at distance $L$ is
$k_BT\ln(c_Lv_0)$).  The potential $\psi(L)=ZeQ^*/DL$ obviously does
not vanish at distance $L$.

Correspondingly, Eq. (\ref{equi2}) is replaced by:
\begin{eqnarray}
Ze[\psi(0)-\psi(L)]&=&|\mu_{WC}|-k_BT\ln\frac{1}{\alpha(A-a)^2\Lambda
  c_L}- \nonumber \\
 &&~~~ -k_BT\frac{a}{A-a}~.
\label{equi2L}
\end{eqnarray}
In the simulation, the $Z$-ion concentration $c_L$ (hence, their
chemical potential) is kept relatively
constant by choosing the number of $Z$-ions at any $Q$ in such a way
that after neutralizing the macroion charge, there are 55 $Z$-ions
left. As one can see from Fig. \ref{mcresult},
the oscillations of the number of $Z$-ions adsorbed at the
macroion surface do not exceed 4. 
Such a small deviation in the number of free $Z$-ions 
from 55 only weakly affects $c_L$. 
Furthermore, due to the excess charge of these extra $Z$-ions, they
spread out in a thin layer near the outer boundary where their charges
are neutralized. This helps to keep $c_L$ relatively constant.
Because in Eq.
(\ref{equi2L}), $c_L$ appears under the logarithmic function,
the chemical potential of $Z$-ions oscillates even weaker
than $c_L$.

The solution to Eq. (\ref{equi2L}) is plotted by the solid line in
Fig. \ref{mcresult}.  As we can see, although the theoretical
prediction agrees favourable with the Monte Carlo results, it
overestimates the net inverted charge near the full filling.  This
discrepancy is due to the use of a planar hard disc system in the
theory of the $Z$-ion layer at the macroion surface.  Indeed,
unavoidable lattice defects caused by the curvature of the macroion
surface (there are 12 disclinations and a number of dislocations in
the $Z$-ion layer\cite{Koulakov}) can lead to a large misprediction of
the number of $Z$-ions at complete filling if we assume a flat layer
with the same area.  For example, calculations using icosahedron show
that even in the optimal situations with ``magic'' numbers of
$Z$-ions, planar geometry mispredicts the number of $Z$-ion at
complete filling by 1 for $N=12$ and by 3 for $N=32$. Away from the
magic numbers where a number of dislocations appear, this
misprediction is even larger.  The dash line in Fig. \ref{mcresult} is
the adjusted theoretical prediction of Eq. (\ref{equi2L}) assuming the
number of $Z$-ions at maximum filling of the first layer is 79 instead
86. The agreement between this curve and the Monte Carlo results is
much better than the solid curve. It should be noted that, defects are
important only near complete filling where the $Z$-ions layer forms a
real solid. At smaller filling where $Z$-ion layer is liquid like,
defects play no role. That is why the adjusted prediction (the dash
line in Fig. \ref{mcresult}) is not drawn at smaller filling.
Obviously, at larger macroion size, 
the region where defects are important is
relatively smaller.

The planar approximation of the macroion surface also forces us not to
extend theoretical prediction to $Q<48$ (or 12 $Z$-ions) because, 
at small $Q$, the average distance between
$Z$-ions on the macroion surface $2A=\sqrt{8\pi(r_M+a)^2Ze/\sqrt{3}Q}$
is larger than the radius $r_M$ of the macroion. This makes the
concept of a  two-dimensional $Z$-ion layer, and therefore Eq.
(\ref{equi2L}), quantitatively incorrect. Again, if the
macroion size gets larger or the $Z$-ion size gets smaller, the
$Z$-ion layer can be truthfully regarded as two-dimensional in a
larger region of $Q$.  This in turn makes Eq.  (\ref{equi2L})
applicable quantitatively in a wider region of $Q$.  It should be
noted that, even though Eq. (\ref{equi2L}) fails at small $Q$ in our
case, the qualitative picture of charge inversion is still valid. This
is because, in our case, the plasma parameter $\Gamma$ is still large
and does not reach unity until $Q\simeq 2$.

\section{Voltage controlled oscillations of charge and capacitance}
\begin{figure}
\epsfxsize=8.5cm \centerline{\epsfbox{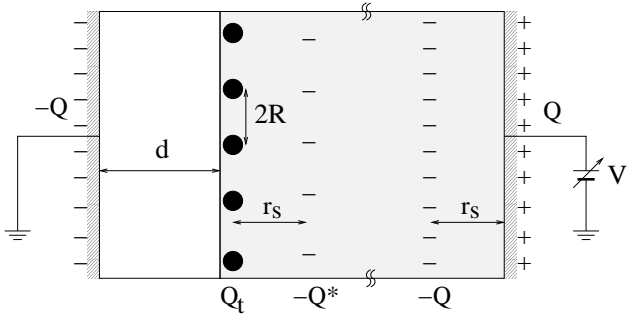}}

\vspace{.3cm}

\caption{A sketch of the proposed experiment. Two metallic plates are separated
  by an insulator (white) and a water solution (gray) containing
  multivalent $Z$-ions (shown by solid circles). Under the voltage $V$
  applied to the capacitor, charges $-Q$ and $Q$ are induced in the
  metallic plates. $Z$-ions with total charge $Q_t=Q^*+Q$ are
  adsorbed on the right side of the insulator. The excess charge $Q^*$
  due to charge inversion is linearly screened at the distance $r_s$ by
  monovalent salt.}
\label{exp}
\end{figure}

In the previous sections, we discussed the oscillating behaviour of
the inverted charge as a function of the macroion charge.  Generally
speaking tuning the charge of an insulating 
macroion is not a trivial task, although can
be done \cite{Cuvillier}. We propose here an alternative experiment
that would show this oscillating behaviour when the macroion surface
potential is tuned instead. It is sketched in Fig.  \ref{exp}.
The role of the macroion is played by the left metallic plate covered by an
insulator layer with thickness $d$, surface area $S$ and dielectric
constant $D_i$.  The other side of the dielectric is in contact with
the ionic solution which contains the $Z$-ions. Some of $Z$-ions
condense on the dielectric surface to screen the electric field
created by the negative charges on the metal plate. By changing the
voltage $V$ applied to the other plate, one can easily vary the charge
of condensed $Z$-ions on the dielectric surface.  We show below that
if one measures the differential capacitance of this systems as a
function of $V$, one can expect an oscillating behaviour.

To deal with a macroion at a constant potential, our theory needs some
modifications.  When the voltage $V$ is applied, a charge $-Q$ is
induced on the left metallic plate. $Z$-ions layer with total charge
$Q_t$ is then formed on the dielectric surface in solution to screen
the electric field of the plate. The excess charge $Q^*=Q_t-Q$ due to
overscreening is in turn linearly screened at the distance $r_s$ in the
solution. Due to the electro-neutrality of the solution, charge $-Q$
appears on the other side of the solution container to screen the
second metallic plate. If $d$ is large enough ($d \gg A$) so that one
can neglect the image of $Z$-ions in the left metallic plate, the free
energy of the system can be written as:
\begin{equation}
F=\frac{Q^2}{2C(d)}+\frac{Q^{*\,2}}{2C(r_s)}+N\varepsilon(n)+
        \frac{Q^2}{2C(r_s)}-QV-N\mu_b~,
\label{freeC}
\end{equation}
where $Q$ is the charge, $C(d)=D_iS/4\pi d$ is the capacitance and
$Q^2/2C(d)$ is the energy of
the planar capacitor formed by the dielectric layer with thickness
$d$. As we mentioned above, the excess charge $Q^*$  is linearly 
screened at distance $r_s$ in solution. The free energy of this
screening atmosphere can be expressed as the energy of a capacitor
with charge $Q^*$, thickness $r_s$ and capacitance $C(r_s)=DS/4\pi
r_s$ (The second term in Eq. (\ref{freeC})).  The third term in
Eq. (\ref{freeC}) accounts for the 
correlation energy of the SCL of $Z$-ions condensed at the dielectric
surface ($N=Q_t/Ze$ is the number of $Z$-ions condensed and $n=N/S$ is
their two-dimensional concentration).  The fourth term in Eq.
(\ref{freeC}) is the energy of the capacitor with thickness $r_s$ on
the right side of the solution container. On this side only monovalent
ions screen the electrode.  Therefore correlations are weak and
screening can be described by the usual linear theory which gives the
free energy $Q^2/2C(r_s)$.  The fifth term in Eq. (\ref{freeC}),
$-QV$, is the work the source does to maintain the system at the
constant voltage $V$. Finally, the last term in Eq. (\ref{freeC}) is
the change in the free energy of the bulk solution which acts as a
reservoir to provide the $Z$-ions to the macroion surface.

Minimizing the free energy (\ref{freeC}) with respect to the unknown
parameters $Q$ and $Q_t$, one gets:
\begin{eqnarray}
&&V=Q\left(C(d)^{-1}+2C(r_s)^{-1}\right)-Q_tC(r_s)^{-1} \label{eqV}~,\\
&&-(\mu(n)-\mu_b)/Ze=Q^*C(r_s)^{-1}~.
\label{eqQ}
\end{eqnarray}
As shown in previous sections, due to correlations between $Z$-ions,
$-(\mu(n)-\mu_b)$ is positive everywhere except near complete filling.
According to Eq. (\ref{eqQ}), this means that $Q^*$ is positive,
indicating charge inversion.

Solving Eqs. (\ref{eqV}) and (\ref{eqQ}), one gets
\begin{equation}
V=Q\left(C(d)^{-1}+C(r_s)^{-1}\right)+\frac{\mu(n)-\mu_b}{Ze}~~,
\end{equation}
which shows that the correlation between $Z$-ions condensed on the
dielectric surface works as an additional voltage applied to the
system.
 
A measurable quantity of the system is its differential capacitance
$C=\partial Q/\partial V$, or
\begin{equation}
C^{-1}=\frac{\partial V}{\partial Q}=C(d)^{-1}+C(r_s)^{-1}
        +\frac{1}{Ze}\frac{\partial \mu(n)}{\partial n}~~.
\label{eqCinv}
\end{equation}
As we saw in previous sections, excluded volume effect leads to the
oscillations in the chemical potential $\mu(n)$ of the $Z$-ions which
becomes positive every time a full layer of $Z$-ions is developed.
From Eq. (\ref{eqCinv}), this ultimately results in the oscillations
of the inverse capacitance $C^{-1}$ of the system. Thus measuring
$C(V)$ one can indirectly observe the oscillations in charge
inversion.

Concluding this section, we would like to mention that the
oscillations of the differential capacitance we obtained above are
remarkably similar to the well known oscillations in the
compressibility and magneto-capacitance in the two-dimensional
electron gas in quantum Hall regime~\cite{Smith}.
In this case, the oscillations are related to the consecutive
filling of Landau levels. The compressibility of the two-dimensional
electron gas vanishes every time a Landau level is fully occupied. In
our system, the two-dimensional  electron gas is replaced
by the quasi two-dimensional liquid of $Z$-ions and the magnetic field
is replaced by the varying voltage. In this sense, we can say that we
are dealing with a classical analog of the quantum Hall effect.

\section{Conclusion}
Summarizing our results, we have shown that the excluded volume of
$Z$-ions can dramatically change the screening of a macroion by
$Z$-ions.  In particular, charge inversion is shown to oscillate with
the macroion surface charge density $\sigma$. At small macroion
surface charge density, it increases with $\sigma$. However, when the
surface charge is high enough so that a near full layer of $Z$-ion is
formed, charge inversion disappears quickly because of the vanishing
compressibility of a full layer of $Z$-ions. Charge inversion recovers
when $\sigma$  increases further so that a second layer of $Z$-ions
builds up. Since tuning of the charge of an insulating 
macroion is not a simple task,
we also propose an experimental setup with metallic electrodes playing
the role of a macroion. This experiment can be used to indirectly
observe these oscillations of charge inversion.

Oscillations of charge inversion discussed in this paper are related
to the hard-core repulsion of spherical $Z$-ions. Similar effect
should exist for the adsorption of rigid rod-like polyelectrolytes
(PE) on oppositely charged surface.  
On a strongly charged surface rods can form more
than one layer leading to a deep minimum of the charge inversion every
time when the top layer is completely full.  It is important that
 charge inversion in both cases is determined by the top layer.
 
It is interesting to discuss the relation between our work and the
recent theory of adsorption of a thick layer of weakly charged,
flexible PE with excluded volume~\cite{Rubinstein} in which no
oscillations of charge inversion with the increasing surface charge
density of macroion, $\sigma$, was found.  Instead, the inverted
charge reaches a constant maximum value when PE molecules start to
overlap.  The picture described in Ref. \onlinecite{Rubinstein} is as
follows.  A flexible PE molecule in solution can be viewed as a
rod-like string of Gaussian electrostatic blobs.  Each such blob has a
quite low density of polymer. At small absolute value of $\sigma$, PE
rods screen the macroion surface forming correlated liquid of locally
parallel rods which overcharges the macroion.  As $\sigma$ increases,
charge inversion also increases until, at some value of $\sigma$
which the authors call
$\sigma_e$, the blobs start overlapping with
each other. In contrast with the hard-core repulsion case, blobs are
easily compressible. Therefore, when the surface charge $\sigma$
increases further, the adsorbed PE molecules form a scaling 
structure of layers of blobs with increasing size and decreasing density from
the bottom to the top. In the top layer, blobs have the same size as
unperturbed blobs in the solution.  As in the case of hard core
compact $Z$-ions, it is this top layer which is responsible for the
image of an
approaching new PE molecule and, therefore, for the correlation induced
charge inversion. According to Ref.  \onlinecite{Rubinstein}, in the
case of flexible weakly charged PE the properties of the top layer do
not oscillate with $\sigma$ because newly adsorbed PE molecules lead
to an additional compression of lower layers. Thus at $\sigma >
\sigma_e$, the inverted charge does not oscillate and remains the same
as at $\sigma = \sigma_e$.
 
\acknowledgements
 
We are grateful to A. Yu. Grosberg, V. Lobaskin, W. Halley and M.
Rubinstein for useful discussions.  This work was supported by NSF
DMR-9985785.

\section{appendix}
To derive the chemical potential (\ref{mu}), let us start with the
standard relationship 
between the free energy and the canonical partition function.
\begin{equation}
F(N,V,T)=-k_BT\ln {\cal Z}
\label{F}
\end{equation}
where the canonical partition function $\cal Z$ is given by:
\begin{eqnarray}
{\cal Z}&=&\frac{1}{s_0^{N}N!}\int d{\bbox r}_1...d{\bbox r}_N 
        ~\exp\left\{-\frac{1}{k_BT}\left[\sum_{i=1}^{N}Zev_e({\bbox r}_i)
                 +\right.\right. \nonumber \\
        && \left. \left.
                \sum_{i<j=1}^{N}
                        \left[v_C(|{\bbox r}_i-{\bbox r}_j|)
                +v_{hc}(|{\bbox r}_i-{\bbox r}_j|)\right] \right]
        \right\}~~.
\label{Z}
\end{eqnarray}
Here $v_e({\bbox r}_i)$ is the electrostatic external potential
at the position of the particle $i$ (in our case, $v_e(\bbox r)$ is
the potential due to the macroion charge), $v_C(r)=Z^2e^2/Dr$ is the
usual Coulomb part of the pair potential and $v_{hc}(r)$ is the
hard-core repulsion part:
\begin{equation}
v_{hc}(r)=\left\{
               \begin{array}{ll}
                        \infty  &      \mbox{if $r<2a$} \\
                        0 &   \mbox{if $r>2a$}
                \end{array}
        \right.
~~~,
\end{equation}
and $s_0$ is the normalizing area.
In writing down Eq. (\ref{Z}), we use the primitive model of the
solution where all water degrees of freedom are averaged out,
resulting in an effective dielectric constant $D$ and an effective
quantum cell area $s_0$.

Substituting Eq. (\ref{Z}) into (\ref{F}) and differentiating with respect to
the particle charge $e$, one gets
\begin{equation}
\frac{\partial F}{\partial e}=\left\langle\frac{2}{e}
        \left[\sum_{i=1}^{N} v_e({\bbox r_i})+
                \sum_{i<j=1}^N v_C(|{\bbox r}_i-{\bbox r}_j|)\right]
        \right\rangle
\label{Fdiff}
\end{equation}
here the notation $\langle ...\rangle$ denotes the statistical
average.

From Eq. (\ref{Fdiff}), one gets the final expression for the free
energy of the system:
\begin{equation}
F=F_0+\int_0^e\frac{2{\rm d}e}{e}\left\langle
        \sum_{i=1}^{N} v_e({\bbox r_i})+
                \sum_{i<j=1}^N v_C(|{\bbox r}_i-{\bbox r}_j|)
        \right\rangle
\end{equation}
where $F_0$ is the free energy of the systems of hard discs with
radius $a$ at the same concentration. The integration denotes the
electrostatic contribution to the free energy. As in the case of a SCL
of point charges, because $\Gamma$ is large, one can approximate this
electrostatic contribution by that of a Wigner crystal:
\begin{equation}
\label{Ffinal}
F\simeq F_0+N \varepsilon_{WC}(n)~~.
\end{equation}
Due to hard-core repulsion, the $Z$-ions are never at a distance
smaller than $2a$ from each other. Therefore, the approximation made
in Eq. (\ref{Ffinal}) is even better than that of
point charge SCL and is asymptotically exact when $A \rightarrow a$.

Eq. (\ref{mu}) can now be easily obtained by differentiating
Eq. (\ref{Ffinal}) with respect to the number of $Z$-ions $N$.

\end{multicols}

\begin{references}
\bibitem{metallization}It should be noted here that there are
several works which claim to get charge inversion 
in the Poisson-Boltzmann (PB)
approximation\cite{Joanny0,Andelman} for the case of
polyelectrolyte adsorption on oppositely charged surface.
Careful analysis\cite{Nato} shows
that what is done in Ref. \onlinecite{Joanny0,Andelman} is not exactly
PB approximation but another approach which we call metallization.
Metallization smeares adsorbed polyelectrolyte charge
on the macroion surface thus eliminating its self energy.
On the other hand,
correlations which we study here keep polyelectrolyte molecules
away from each other and eliminate this self
energy only partially. Thus,
metallization is an approximate way to take
correlations into account. We would like
to emphasize that the metallization approach
is different from the PB one\cite{Nato}.
PB approximation effectively smears
polyelectrolytes everywhere, both at the macroion surface and in
the bulk of solution, while the metallization
approach keeps polyelectrolytes
of the bulk discrete (which results in their rod like shape there)
and uses the elimination of the self-energy of charged rods at
the macroion surface as the driving force of charge inversion.
Therefore, it is not surprising that
there is no charge inversion in PB approximation, while
the metallization approach even somewhat overestimates it.
\end{references}
\end{document}